\documentclass[conference]{IEEEtran}

\usepackage{booktabs} 

\usepackage{graphicx}
	\usepackage{amsmath}
	\usepackage{amsfonts}
    \usepackage{longtable}
	\usepackage{fancyhdr}
	\usepackage{mdframed}
\mdfdefinestyle{style1}{leftmargin=1cm,rightmargin=0.5cm,skipbelow=0.5cm,skipabove=0.2cm,
   innertopmargin=2pt}
	\usepackage{balance}
    \usepackage{fancybox}
    \usepackage{booktabs}
	\usepackage{enumerate}
    \usepackage{framed}
    \usepackage{pgfplots}
	\usepackage{listings}
    \usepackage{subcaption}
    \usepackage{xcolor}
    \usepackage{hyperref}

\ifCLASSOPTIONcompsoc
  \usepackage[nocompress]{cite}
\else
  \usepackage{cite}
\fi

\begin{document}
\title{NALABS: Detecting Bad Smells in Natural Language Requirements and Test Specifications}

\author{Kostadin Rajkovic, Eduard Enoiu\\
      M\"alardalen University, V\"aster{\aa}s,  Sweden.
}

\IEEEtitleabstractindextext{%
\begin{abstract}

In large-scale embedded system development, requirement and test specifications are often expressed in natural language. In the context of developing such products, requirement review is performed in many cases manually using these specifications as a basis for quality assurance. Low-quality specifications can have expensive consequences during the requirement engineering process. Especially, if feedback loops during requirement engineering are long, leading to artifacts that are not easily maintainable, are hard to understand and inefficient to port to other system variants. We use the idea of smells to specifications expressed in natural language, defining a set of specifications bad smells. We developed a tool called NALABS (NAtural LAnguage Bad Smells), available on \url{https://github.com/eduardenoiu/NALABS} and used for automatically checking specifications. We discuss some of the decisions made for its implementation, and future work.
\end{abstract}
}
%
%



\maketitle
\IEEEdisplaynontitleabstractindextext
\IEEEpeerreviewmaketitle

\section{Introduction}
In many embedded system domains, requirement engineering is performed manually by engineers that are hand-crafting those artifacts using natural or semi-structured languages and using them for other stages of software development, such as testing them against the system-under-test. Issues in natural language artifacts (i.e., requirements), such as ambiguities or vague specifications, can lead to higher costs during system development \cite{fernandez2017naming}. 
Previous research found that test specifications written in natural language contain a rather high degree of cloning and bad structure \cite{hauptmann2012can,hauptmann2013hunting} which can influence the cost of maintaining and executing test cases. For requirements \cite{femmer2017rapid,femmer2017rapid,unterkalmsteiner2017requirements} as well as for system test cases \cite{hauptmann2013hunting} written in natural language, so called bad smells have been established as indicators to identify poorly
written natural language artifacts. 

Based on several natural language smells observed in previous studies, we established a set of indicators for requirement flaws and defined dictionary-based metrics to automatically detect these smells in natural language artifacts. This paper introduces NALABS, a new tool that aims at
quick analysis of requirements by detecting problematic requirements.

\section{Related Work}
Related to quality assurance of other natural language artifacts such as requirements, several researchers have focused on detection of clones \cite{juergens2010can}, requirement similarity \cite{falessi2011empirical} and ambiguity \cite{gleich2010ambiguity}. Femmer et al. \cite{femmer2017rapid} proposed to detect issues based on requirement smells based on the quality attributes of natural language requirements in ISO/IEC/IEEE 29148. Hauptmann et al \cite{hauptmann2013hunting} is the first to study smells in natural language test specification. Hauptmann \cite{hauptmann2016reducing} proposed several natural language
test smells to identify quality problems in test cases as well as showing the results of applying this approach on nine industrial test suites. We identified gaps in the previous research which we are tackling in this paper. Firstly, the set of smells for natural language artifacts used in some previous studies is limited and mainly based on maintainability attributes. We extend this list by tailoring requirement-based bad smells to other metrics related to complexity. In addition, only two studies apply their work on real industrial specifications. The aim of NALABS is to help in bringing more evidence on the industrial use of bad smells for detecting specification quality defects. 

\section{Natural Language Specifications Smells} 

In this section, we describe the quality of natural language specifications, how we created the set of smells and the automatic measurement of these smells using specific dictionaries. 

\begin{itemize}
    \item \textbf{Vagueness} is a common problematic property when it comes to understanding requirements and requirement issues. Many papers are identifying this as an indicator of quality problems, but we have found definitions on how to measure it only in two papers \cite{rendex, quars}. It is interesting that the suggested lists of keywords are completely different in these two papers. However, the list of keywords found in \cite{quars} was incomplete and only the definition found in \cite{rendex} is used in NALABS. The following list of keywords has been used: may, could, has to, have to, might, will, should have + past participle, must have + past participle, all the other, all other, based on, some, appropriate, as a, as an, a minimum, up to, adequate, as applicable, be able to, be capable, but not limited to, capability of, capability to, effective, normal.
     \item \textbf{Referenceability.} Number of reference documents\cite{rendex} was found in a couple of papers under different names (i.e., Directive Frequency \cite{quars}, Directives\cite{thenasa}). This is usually an indication of nesting in the requirements documents or a need for additional reading in order to understand the requirement that contains references. The issue on how to measure this indicator it is not well defined. In \cite{rendex, thenasa} it was suggested to count the overall occurrence of keywords that would indicate referencing. In \cite{quars} the list of keywords was not provided, but the authors suggested that the rate of pointers to figures, tables, notes should be counted. We have decided to implement the measures from \cite{rendex} and \cite{thenasa} as two separate measures (NR1 and NR2). The authors of \cite{rendex} have anticipated the need of adjusting this measure for different writing styles of each company and therefore we have decided to expand the list of keywords by adding the keyword "see", since this is the most common writing style of referencing noticed in requirements documents in a large manufacturing company in Sweden.
\textbf{List of keywords (NR1 \cite{rendex}):}  defined in reference, defined in the reference, specified in reference, specified in the reference, specified by reference, specified by the reference, see reference, see the reference, refer to reference, refer to the reference, further reference, follow reference, follow the reference, see document.
\textbf{List of keywords (NR2 \cite{thenasa}):} For example, Figure, Table, Note.
     \item \textbf{Optionality.} This metric was found in certain papers \cite{SATC} and \cite{quars} with similar definitions. Optional words are giving the developers a latitude of interpretations to satisfy the specified statements and their use is usually not recommended in requirements documentation. Authors have proposed different lists of keywords, but the list found in \cite{quars} was incomplete and therefore we have decided to select the measure found in \cite{thenasa}.
\textbf{List of keywords:} can, may, optionally.
     \item \textbf{Subjectivity.} Subjectivity metric is measuring personal opinions or feelings in sentences. It was proposed in \cite{quars} with a list of keywords to be counted in text. However, since one of the proposed phrases to detect was "as [adjective] as possible", we have decided to avoid detection of adjectives in text, and replace this phrase and detect only the last part of this sentence.
\textbf{List of keywords:} similar, better, similarly, worse, having in mind, take into account, take into consideration, as possible
     \item \textbf{Weakness} \cite{quars} \cite{thenasa} is a metric that counts words and phrases that may introduce uncertainty into requirements statements by leaving room for multiple interpretations. The metric was found in two mentioned papers, but we have decided to use only the one found in \cite{thenasa}, since the list of keywords found in \cite{quars} was very similar to the keywords used for the Optionality metric.
\textbf{List of keywords:} adequate, as appropriate, be able to, be capable of, capability of, capability to, effective, as required, normal, provide for, timely, easy to.
     \item \textbf{Readability.} Automated Readability Index (\textbf{ARI}) is calculated using $WS+9 \times SW$, where WS is the average number of words per sentence and SW is the average number of letters per word \cite{quars}. Readability is considered by a couple of other papers as well, mostly by the use of different readability indexes such as Flesch reading ease index, Flesch-Kincaid, Coleman-Liau and Bormuth grade level index. We have decided to use ARI for its simplicity of implementation.   
     \item \textbf{Complexity Metrics}. Measuring the size of the requirement was used in a couple of discovered papers and it is defined differently in different papers \cite{rendex,thenasa}. It can be defined in many different ways such as the total number of characters, number of words, paragraphs and lines of text. We have decided to use the number of words as the measure of size to account for different writing styles. In addition, we count the overall number of occurrences of conjunctions. This measure was found to be a context-independent measure and can show relations and actions.
\end{itemize}

\begin{figure}[t]
  \begin{center}
    \includegraphics[width=\linewidth]{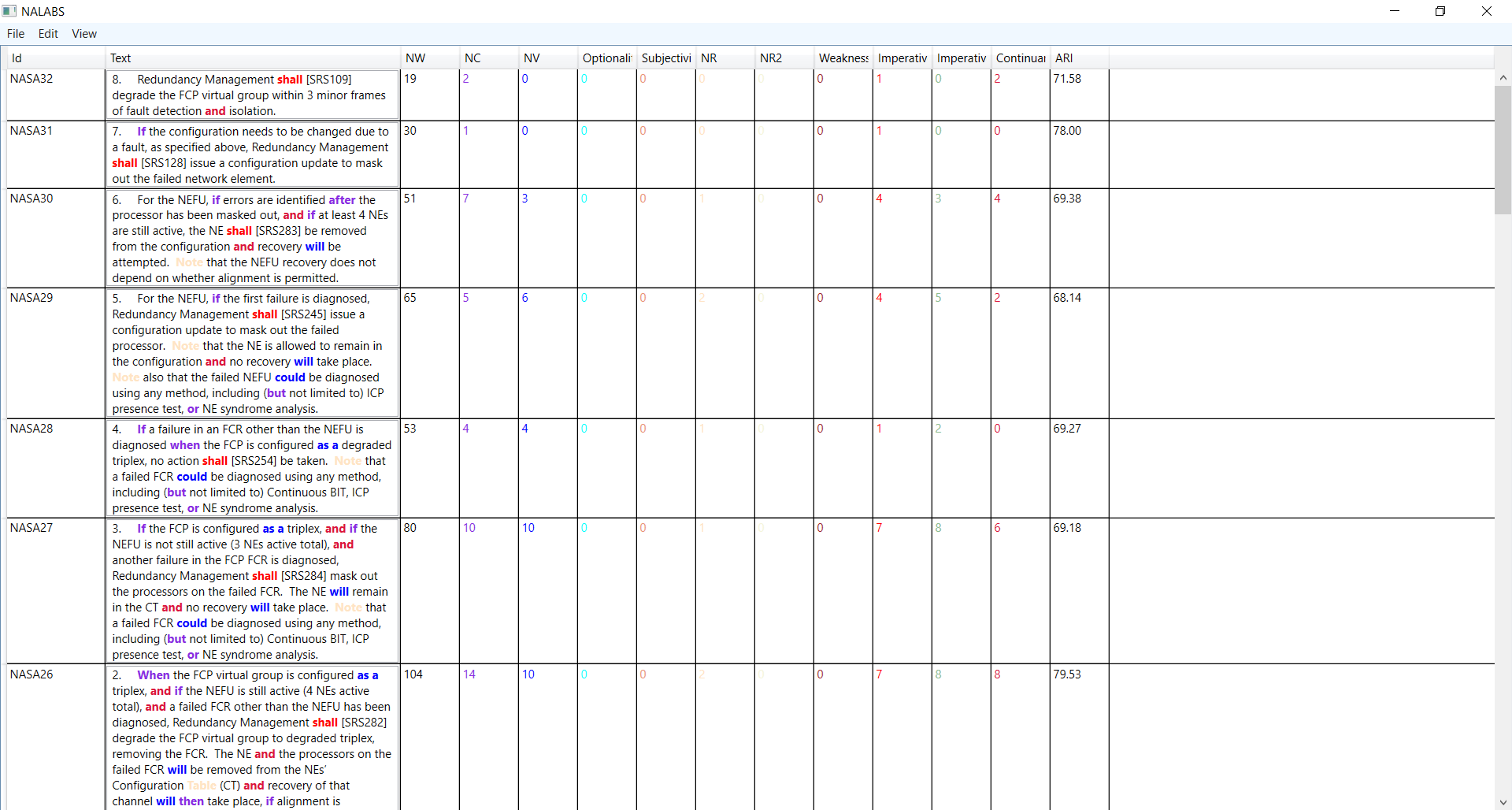}
    \caption{
      Usage of NALABS from GUI.
      \label{fig-context} }
  \end{center}
\end{figure}

\section{Tool Description and Future Work}
Figure \ref{fig-context} shows the use of NALABS from a GUI interface. Although NALABS is still in an early phase of development (it was started in the late 2019), it has already been
used to successfully find several problems in requirements in existing industrial projects \cite{rajkovic2019measuring}. NALABS is
released under the MIT open-source license, and it is freely
accessible on GitHub\footnote{https://github.com/eduardenoiu/NALABS}. It is a desktop WPF application that depends on standard .NET packages. The tool was developed in C\# as a desktop
application for Windows OS and contains three essential layers: (i) the pre-processing of requirement documents stored as excel spreadsheets\footnote{Many companies and tools (such as IBM's Rational DOORS) use this format for working with requirements.}, (ii) the configuration and application of bad smells metrics and (iii) presenting these results to the user.  Future research should focus on exploring and proposing new bad smells measures, combining the existing measures in a single index
of quality and complexity as well as exploring new ways of applying such requirement checkers in industrial systems.


\section{Acknowledgements}
NALABS has been funded by Bombardier Transportation through a thesis project, by the European Union’s Horizon 2020 research and innovation program under grant agreement No. 957212 and by the Swedish Innovation Agency (Vinnova) through the XIVT project. This work was partially funded from the Electronic Component Systems for European Leadership Joint Undertaking under grant agreement No. 737494 and The Swedish Innovation Agency, Vinnova (MegaM@Rt2).

\balance
\bibliographystyle{IEEEtran}
\bibliography{acmart} 

\end{document}